\documentclass[aps,prl,showpacs,twocolumn,superscriptaddress]{revtex4}
 \usepackage{amsmath}
\usepackage{graphicx}
\usepackage{epsfig}
\newcommand{\beq}{\begin{equation}}
\newcommand{\eeq}{\end{equation}}
\newcommand{\bea}{\begin{eqnarray}}
\newcommand{\eea}{\end{eqnarray}}

\newcommand{\sx}{\sigma_{\rm x}}
\newcommand{\sy}{\sigma_{\rm y}}
\newcommand{\sz}{\sigma_{\rm z}}
\begin{document}
\bibliographystyle{apsrev}
 
\title{ Scaling and interaction-assisted transport  in  graphene with one-dimensional defects}
\author{  M. Kindermann}
\affiliation{ School of Physics, Georgia Institute of Technology, Atlanta, Georgia 30332, USA  }

\date{ March 2010}
\begin{abstract} 
  We analyze the  scattering from one-dimensional  defects  in intrinsic graphene. The  Coulomb repulsion between electrons  is found to be able to induce singularities of such scattering    at zero temperature   as in one-dimensional conductors.  In striking contrast to electrons in one space dimension, however, repulsive interactions here can  {\em enhance}  transport. We present explicit calculations for the scattering from vector potentials that appear when strips of the material are under strain. There the predicted effects are  exponentially large for strong scatterers. 
\end{abstract}

\maketitle
 The Coulomb repulsion between electrons can have profound consequences for the scattering of electrons in conductors. In one-dimensional (1D) conductors  it  dramatically suppresses  the conductance through impurities,  with a singularity at   temperature $T=0$.  This is one of the hallmarks of the   Luttinger liquid state of interacting electrons in 1D     \cite{kane:prb92}.  A particularly intuitive understanding of this effect is due to Matveev, Yue, and Glazman     \cite{matveev:prl93,yue:prb94}. In their approach     interactions cause extra electron  scattering from  Friedel oscillations, which inhibits   transport. Friedel oscillations are density modulations that form when  electron waves incident on an impurity interfere with backscattered waves. In 1D  at $T=0$ the  Friedel oscillation around an impurity   is not integrable over space and the induced scattering amplitude diverges  logarithmically in Born approximation.  
 
 Friedel oscillations in  two   dimensions (2D) at nonzero chemical potential can be integrated  and have much weaker effects for both, point-like  \cite{altshuler:bo85, zala:prb01,cheianov:prl06} and 1D defects \cite{shekhtman:prb95,alekseev:prb98}.  
 Interactions  play a more important role in    graphene \cite{novoselov:sci04,zhang:nat05,berger:jpc04} at zero chemical potential (``intrinsic graphene''), which forms a so-called marginal Fermi liquid 
 \cite{gonzalez:prb99}.
 There is, for instance,  experimental evidence \cite{li:nph08} of a singular renormalization of the Fermi velocity  \cite{gonzalez:prb99,barlas:prl07}.  Nevertheless, also  in intrinsic graphene scattering from point-like impurities  of both, scalar and vector potential character,  does not receive any singular corrections from the Coulomb  interaction    \cite{stauber:prb05,foster:prb08} \footnote{Staggered potentials do have singular interaction corrections \cite{stauber:prb05,foster:prb08}, but they require atomically sharp disorder.}.
 
  In this Letter we show  that the Coulomb interaction  in intrinsic graphene {\em can} have a singular impact on  scatterers that are extended in one space dimension. A qualitative analysis of the Friedel oscillations at such scatterers reveals the reason: the Friedel oscillation due to an electron with wavevector $k'$ scattering with amplitude $r$ from a  1D defect  fills a strip of width $\sim1/k'$. The corresponding exchange potential has support in the same strip, implying a factor $\sim r/k'$ in the induced Born scattering amplitude $\delta r^{\rm ex}$. A second factor $1/k'$ comes from the Coulomb interaction. For  energy-independent $r$ the exchange interaction with electrons in  all filled  states thus results in   $\delta r^{\rm ex}\propto r \int_{\rm filled} d^2k'/k'^2$. Since   in intrinsic graphene at $T=0$ the entire valence band is filled,   $\delta r^{\rm ex}$ is  logarithmically divergent. 
This suggests that when the Coulomb interaction  does not  cause a Hartree potential it can have  the same drastic effects for  1D scatterers in 2D intrinsic graphene      as for point defects in 1D conductors. 

In contrast to the familiar situation in 1D  \cite{kane:prb92}, however, we demonstrate that    the Coulomb interaction in graphene can  {\em enhance} electron transport. This intriguing effect is due to bound states that  straddle defects. The exchange interaction of transport electrons with electrons   in such bound states opens an additional transport path across the defect, which increases the electric current.

  \begin{figure}[h]
\includegraphics[width=8cm]{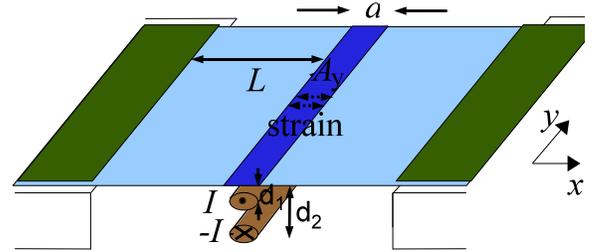}
\caption{A vector potential $A_y$ in a narrow strip (dark blue)  forms a scattering barrier in a sheet of otherwise ballistic graphene (light blue with contacts in green). $A_y$ may be induced by strain   or, alternatively, by a pair of current-carrying wires (brown).}  \label{fig1} 
\end{figure}

We exemplify the above effects with the scattering barrier shown in Fig.\ \ref{fig1},  formed by a vector potential  in a narrow strip of an otherwise ballistic sheet of intrinsic graphene. Such vector potentials are induced   by strain  \cite{fogler:prl08,pereira:prl09} or by nearby electric currents, see Fig.\ \ref{fig1}. By symmetry vector potentials in graphene do not produce a Hartree potential. As suggested by the above argument, the Coulomb interaction therefore induces extra scattering whose  amplitude diverges at low $T$, when the electron wavelength is much longer than the width $a$ of the barrier and scattering is energy-independent. The  effect  turns out to be exponentially enhanced for strong barriers, where it strongly impacts transport  even at moderate $T$.  
 In addition, the scatterer of Fig.\ \ref{fig1} hosts bound states that enhance transport through  electron exchange as motivated above.  As a consequence,    the barrier of  Fig.\ \ref{fig1}  becomes entirely transparent at $T=0$, in striking contrast to impurities in interacting 1D conductors that completely block the electric current at $T=0$. 

 {\em Model:} In the setup of Fig.\ \ref{fig1}  a vector potential
  \beq \label{vec}
 {\bf A}=\left(0,A_y\right);\;\;A_y=\frac{\tilde{\chi}}{ea}\left[\Theta\left(x+\frac{a}{2}\right)-\Theta\left(x-\frac{a}{2}\right)\right]
 \eeq
[$\Theta(x)$ is the step function] is applied to  a  sheet of clean, intrinsic graphene  that we first assume to be infinite, $L\to\infty$. The noninteracting  Hamiltonian in valley $\gamma$ thus is
\beq \label{Dirac}
H_\gamma= v \boldsymbol{\sigma_\gamma} \cdot \left(\boldsymbol{p}-e\boldsymbol{A }\right).
\eeq
Here, $\boldsymbol{ \sigma}_\gamma=(\sigma_x,\gamma\sigma_y)$ is a vector of Pauli matrices,  $\boldsymbol{p}$  is the electron momentum,  $v$ the electron velocity, and $-e$  the electron charge.  The above model has parity symmetry, ${\cal P} H{\cal P}=H$, where ${\cal P} \psi(x,y)= \sy \psi(-x,y)$, and particle-hole (PH) symmetry, $\sigma_z H\sigma_z=-H$. In intrinsic graphene   both symmetries are also respected   by electron-electron interactions  \footnote{In the case of PH symmetry this is seen easiest in second quantized form, when $H$ is invariant under the transformation $c\to\sigma_z c^\dag$, where $c$ is the electron annihilation operator. PH-symmetry requires vanishing chemical potential and       interaction with the ion charge density.}. Without restriction we assume  $\tilde{\chi}>0$. Transport through the barrier at $\tilde{\chi}<0$ is obtained through reflection on the $x$-axis.  
The above model has been analyzed in Ref.\   \cite{fogler:prl08} on a noninteracting level, where     ${\bf A}$    suppresses linear transport  exponentially in $\tilde{\chi}$ if $kT \ll \tilde{\chi}v/a$ (we set $\hbar=1$). 
 Below we study the effects of the Coulomb  repulsion between electrons on this transport problem for  short barriers ($\Lambda a\gg 1$, but  $\ln \Lambda a \gtrsim 1$, where $\Lambda$ is  the bandwidth of graphene).
 
  We assume  that the graphene sheet is suspended or lying on an insulating substrate, leaving the  Coulomb interaction unscreened with interaction potential $V_{\rm C}({\bf q})= 2\pi \tilde{r}_{\rm s} v/|\boldsymbol{ q}|$ at interaction parameter $\tilde{r}_{\rm s}$ and wavevector ${\bf q}$ \footnote{Screening by the electrons in the graphene sheet itself is negligible at the wavevectors $k'\gg kT/v$   that make the dominant contributions to the analyzed effects \cite{hwang:prb07,wunsch:njp06}.}.   In the weakly interacting limit $\tilde{r}_{\rm s}\ll 1$ that we take henceforth the main effect of electron-electron interactions is   additional elastic scattering, as in the thoroughly studied 1D problem   \cite{matveev:prl93,yue:prb94,fisher:mes96}.  Also like in 1D, such scattering is induced by the Hartree-Fock potential $V^{\rm HF}$ due to  Friedel oscillations at the barrier.   In addition,  ${\bf A}$ here  induces bound states with  an extra contribution to $V^{\rm HF}$.
 
 The potential $V^{\rm HF}$ is strongly constrained by the symmetries of the setup.
Most importantly, a  Hartree potential is precluded by  PH-symmetry. 
Also, the exchange potential $V^{{\rm ex}}$ is spin- and valley-diagonal. Transport in both valleys is   identical   since they are related as  $H_{-1}=U_{\rm v}^\dag H_1U_{\rm v}$, where $U_{\rm v}=\sx$ commutes with the  current $\sx$ through the barrier  \footnote{In case ${\bf A}$ is due to strain,  $\tilde{\chi}$ has opposite sign in the two valleys \cite{fogler:prl08}. Because of the invariance of transport   under $\tilde{\chi}\to-\tilde{\chi}$, however, also this does not affect our results.}. We thus set   $\gamma=1$.

 The  compound scatterer composed of ${\bf A}$ and  $V^{\rm ex}$ is   conveniently described by the transfer matrix of electrons at zero energy $M= T(\infty,-\infty)|_{\varepsilon=0}$ \cite{cheianov:prl06}.   The ${\cal P}$- and PH-symmetries of our model   imply $\sy M\sy=M^{-1}$ and $\sz M\sz=M$, respectively.  
 Together with current conservation $\sx M^\dag\sx=M^{-1}$  they constrain $M$ to take the form $M=\exp(\chi \sz)$.  Low energy transport   in the setup of Fig.\ \ref{fig1}  with weak many-body interactions is thus characterized by a single parameter $\chi$. This reduces our analysis   to a calculation of $\chi$.  At $\tilde{r}_{\rm s}=0$ we have $\chi=\tilde{\chi}$.

{\em First order in $\tilde{r}_{\rm s}$:}   To begin with, we calculate the interaction correction to $\tilde{\chi}$ at first order in $\tilde{r}_{\rm s}$. 
Computing the correction to $M$ due to scattering from the exchange potential created by the noninteracting electron states in Born approximation we find in the limit $a\ll v/kT$ 
\beq \label{dchi}
 \chi =\tilde{\chi}  - \tilde{r}_{\rm s}  \sinh\tilde{\chi} F(\tilde{\chi})\left[\ln \frac{v}{kT a}-\tilde{\chi}+{\cal O}(1)\right] \eeq
with the  positive function
\begin{widetext}
\bea
F(\chi)=\frac{\coth\chi (\cosh3 \chi-3 \cosh\chi) \arcsin(\tanh\chi) +2\cosh2\chi }{2 \pi} - \cosh\chi \sinh^2\chi .
\eea
\end{widetext}

The first order result   Eq.\ (\ref{dchi}) manifests itself in a non-analytic temperature-dependence of the conductance through the barrier, which takes the form
\beq \label{cond}
G=\frac{4e^2}{ h} \frac{kTW\ln 2}{\pi v}\left(1+ \sinh\chi \tanh\chi \,\ln\tanh\frac{\chi}{2}\right).
\eeq
   In $G$, the logarithmic temperature dependence of the interaction correction to $\chi$ competes with  the factor $T$  due to the linear density of states. The predicted non-analytic scale-dependence of  $\chi$  is thus more easily observed in the normalized conductance $G_{\rm r}=G/(G|_{\tilde{\chi}=0})$ \footnote{$G_{\rm r}$ is measured best when ${\bf A}$ is not due to strain, but electric currents that can easily be switched on and off.}.

 Due to its logarithmic $T$-dependence the first order correction to $\tilde{\chi}$ may  grow large  even at $\tilde{r}_{\rm s}\ll 1$. At the corresponding low temperatures, when $\tilde{r}_{\rm s} \ln (v/kTa)\gtrsim 1$, the above perturbative calculation looses validity, as in the 1D case  \cite{matveev:prl93,yue:prb94}. The origin of the logarithmic divergence here is, however, different.  In 1D the divergence stems from a Friedel oscillation density $\propto 1/x$  at $T=0$.  In the present 2D case the $T=0$ Friedel oscillation density is  $\propto 1/x^2$ and integrable. But the non-locality of $V^{\rm ex}$   gives rise to the same  logarithmic divergencies as in 1D.

{\em RG-analysis:} We extend   our analysis into the regime $\tilde{r}_{\rm s}\ll 1$, but $\tilde{r}_{\rm s} \ln (v/kTa) \gtrsim 1$,   by a renormalization group (RG) calculation that  re-sums   the perturbation series in the ``leading logarithm'' approximation \cite{matveev:prl93,yue:prb94}. To this end we successively  integrate out shells of  wavevectors  $[k',bk']$ with $b>1$ in the order of decreasing $k'\ll 1/a$. For each shell we compute the exchange potential $V_{k',b}^{\rm ex}$ due to the corresponding states.  Thereafter we renormalize the transfer matrix for low energy electrons at wavevector $k$ by $V_{k',b}^{\rm ex}$. The resulting  $\chi$  characterizes the wavefunctions in the subsequent shell $[k,bk]$. The logarithmic scale-dependence of  $\delta \chi$, Eq.\ (\ref{dchi}), allows us to assume $k\ll k'$ in this process.  Introducing  $l=-\ln k'a$  we thus find that 
\beq \label{RG}
\frac{d\chi}{dl}= -r_{\rm s}\sinh\chi F(\chi),
\eeq
where $\chi=\tilde{\chi} $ at $l=0$. Here, $r_{\rm s}$ is a scale-dependent interaction parameter that is   renormalized as $  dr_{\rm s}/dl= -r^2_{\rm s}/4$ \cite{gonzalez:prb99} with the solution $r_{\rm s}=\tilde{r}_{\rm s}/(1+\tilde{r}_{\rm s}l/4)$. 
Eq.\ (\ref{RG}) with this $l$-dependent  $r_{\rm s}$  is solved best  by introducing $y=\ln(1+\tilde{r}_{\rm s}l/4)$, such that  $d\chi/dy=-4 \sinh\chi F(\chi)$.

We start the detailed analysis of  Eq.\ (\ref{RG}) assuming  $\tilde{\chi}\gg 1$.  As long as  $\chi\gg 1$ we   use  $\lim_{\chi \to\infty}F(\chi)=2/3\pi$ to find 
\beq \label{exponential}
\frac{d\chi}{dy}=-\frac{4}{3\pi} e^\chi.
\eeq
 Eqs.\ (\ref{RG}) and (\ref{exponential}) make two remarkable predictions: First, the minus signs in both equations imply that  the Coulomb interaction reduces the barrier strength. It thus  {\em enhances} transport through the  studied scatterer,  in evident contrast with the familiar situation in  1D \cite{kane:prb92,matveev:prl93, yue:prb94}. Second, that enhancement is exponentially large in $\chi$ such that for sufficiently large $\tilde{\chi}$ the Coulomb interaction   strongly impacts   electron transport even at $r_{\rm s}l\ll 1$.
  
Both predictions have their origin in bound states. We find that the interaction with the Friedel oscillations due to   propagating states  decreases the conductance as in 1D. That suppression, however, is overcompensated by bound states  which form inside the scattering barrier at  $|x|<a/2$, but leak into the region $|x|>a/2$. Electrons in bound states  have    the same probability for being  to the left of the barrier as to the right, guaranteed by  ${\cal P}$-symmetry  \footnote{The bound states are non-degenerate (apart from the spin and valley degeneracies).}. The  bound states  therefore straddle the barrier and the exchange interaction with electrons  in  those bound states opens an additional   path for  electrons to traverse the barrier. This explains the minus signs in Eqs.\  (\ref{RG}) and (\ref{exponential}). The resulting enhancement of the barrier transmission $t$ is of order $r_{\rm s}l$ and can easily exceed   the exponentially small  noninteracting    $t\propto \exp(-\tilde{\chi})$, which implies the exponential   renormalization of $\chi$, Eq.\ (\ref{exponential}).

\begin{figure}[h]
\includegraphics[width=8cm]{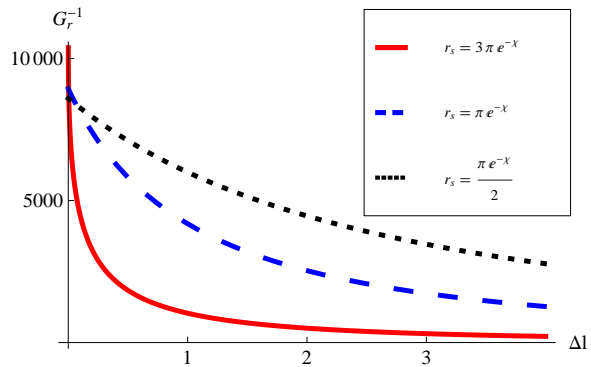}
\caption{Normalized resistance $G^{-1}_{\rm r}$ at  $\tilde{\chi}=5$ and $kT < v\,{\rm sech}\,\tilde{\chi}/a$, when it is renormalized by bound states,
 as  a function of $\Delta l = \ln\left(v\,{\rm sech}\,\tilde{\chi}/kTa\right)$.   }   \label{fig2} 
\end{figure}

This exponential renormalization limits  our RG analysis  to   $\tilde{r}_{\rm s} \ll 3\pi  \exp(-\tilde{\chi})$. 
Also, the bound state energies $\varepsilon^{\rm b}=\pm v k_y {\rm sech}\chi$ are exponentially small at $\chi\gg 1$.  The  exponential renormalization of $\chi$ is thus cut-off at  $\kappa ^{\rm b} =  kT\cosh\chi|_{k'=\kappa^{\rm b}} /v$ and it requires exponentially  low temperatures $kT \ll \exp(-\tilde{\chi})v/a$. 
Eq.\  (\ref{exponential})  is solved by
\beq \label{largechi}
\chi=-\ln\left(e^{-\tilde{\chi}}+\frac{4}{3\pi} y\right),
\eeq
 where $y$ is evaluated at    $\kappa ^{\rm b}$, $y=\ln[1-(\tilde{r}_{\rm s}/4)\ln \kappa^{\rm b}a]$. The self-consistent solution of Eq.\ (\ref{largechi}) with this $\chi$-dependent  $y$ determines the normalized conductance, which for $\chi \gg 1$ takes the form   $G_{\rm r} = 8 \exp(-2\chi)/3$.  In analogy with the Kondo problem we define the temperature scale $T^*$ where the  transmission across the barrier becomes of order unity, $G_{\rm r} \simeq 1$.  At $\tilde{r}_{\rm s} \ll3\pi \exp(-\tilde{\chi})$ we find 
 \beq \label{Ts}
kT^*= \frac{v}{a} e^{-3\pi/\tilde{r}_{\rm s}}.
\eeq
At wavevectors $k'<\kappa ^{\rm b}$  the bound states do not renormalize $\chi$ anymore, but  the propagating states continue to  do so  and  $d\chi/dy=3/2$ for $\kappa^{\rm b} \gg k'\gg kT/v $.   This now {\em decreases} the conductance, by a factor $(1+\tilde{r}_{\rm s}\chi|_{k'=\kappa^{\rm b}}/4)^{-3}$  if  $\chi|_{k'=\kappa ^{\rm b}}\gg 1$. 
In Fig.\ \ref{fig2} we plot   the    normalized resistance $G^{-1}_{\rm r}$  for various parameter values. Some of them are within the regime $\tilde{r}_{\rm s} \ll 3\pi \exp(-\tilde{\chi})$, where  Eq.\    (\ref{largechi}) is rigorously justified, others at  $\tilde{r}_{\rm s} \lesssim 3\pi \exp(-\tilde{\chi})$, where  Eq.\    (\ref{largechi})  may still be qualitatively correct.  At  $\tilde{r}_{\rm s} > 3\pi \exp(-\tilde{\chi})$  Eq.\    (\ref{largechi}) does not have a unique solution anymore.
 Clearly, 
 the predicted effect is large even for moderate $\tilde{\chi} \gtrsim 1$ and   $\ln \kappa^{\rm b} a \gtrsim 1$. Its observation therefore does not require access to large temperature intervals. 

As $\chi$ decreases the scaling behavior crosses over into a second regime, where 
$\chi \ll 1$ and $F(\chi ) $ may be expanded around $\chi =0$. In that case we find $ d\chi/dy=-4(10/3\pi-1) \chi^3$ 
\footnote{This equation has corrections of order $\chi k' a$. It is strictly valid for $k'a\ll \tilde{\chi}$ (e.g. when gates around  $x=0$ screen short wavelength Friedel oscillations), or at   $1/\tilde{\chi}^2 y\ll 1$. } such that, now assuming for simplicity $\tilde{\chi}\ll 1$,
\beq \label{asym}
\chi= \frac{\tilde{\chi}}{\sqrt{1+8\tilde{\chi}^2(10/3\pi-1)y}}.
\eeq
 We conclude that the backscattering from   ${\bf A}$ disappears asymptotically at low $T$, very differently from static scattering potentials $v$ in 1D. The coupling $\chi$, however, is only marginally irrelevant here  ($d\chi/dl\propto \chi^3$ \footnote{The absence of a   term   $\propto \chi$ follows also from Refs.\ \cite{stauber:prb05,foster:prb08}.}). It is  renormalized more slowly than  $v$ in 1D, which  is  relevant in the RG sense ($dv/dl \propto v$) \cite{kane:prb92}. The renormalization of $r_{\rm s}$ in graphene   further slows down the RG flow of $\chi$. 

{\em Discussion:} In reality the unscreened and ballistic segments of graphene around the barrier in Fig.\ \ref{fig1} have a finite length $L$ (typically limited by   contacts   or by the elastic mean free path).   Our  predictions then require  that the   entire Friedel oscillation fits into those segments, implying   an additional cut-off $\kappa^{ L}={\rm coth}\tilde{\chi}/L$ for the RG-flow and necessitating  $L\gg a$. Due to the strong renormalization at $ \chi \gg 1$, however, the observation of the effects shown in Fig.\ \ref{fig2} does not require excessively long samples.

While inelastic processes are negligible in the above analysis of the renormalization of ${\chi}$ at  $r_{\rm s}\ll 1$   \cite{fisher:mes96}, they can  significantly affect the transport between   barrier and   contacts  \cite{fritz:prb08}.  There inelastic scattering can be neglected only if the barrier is more resistive than the adjacent  ballistic regions, such that  $(kT G_{\rm r})^{-1} \gg r_{\rm s}^2 L/v$.  The cut-off $\kappa^{L}$ further limits the regime where  our predictions can be observed directly in the conductance to $r_{\rm s}^{2}G_{\rm r}\ll v/kTL \ll {\rm tanh}\,\tilde{\chi}$. At $r_{\rm s}^{2}G_{\rm r}\gtrsim v/kTL$ the observation of   $G^{-1}_{\rm r}$ requires experimental differentiation from the inelastic background resistance, e.g. by varying $\tilde{\chi}$. 

 The  vector potential ${\bf A}$ is induced by strain in the $x$-direction if that is along the ``armchair'' direction of the graphene lattice \cite{fogler:prl08}. 
Strain, however, 
also generates a scalar, "deformation" potential $V \lesssim \tilde{\chi} v/a$ \cite{fogler:prl08}.  This breaks PH-symmetry, a Hartree-potential appears, and our predictions require $V \ll {\rm max}\{v\coth\tilde{\chi}/L,kT\}/r_{\rm s}$. Also  the bound states are strongly affected by $V$ and consequently   Eq.\ (\ref{exponential}) holds only if $V\ll \tilde{\chi} \exp(-\tilde{\chi})v/a$, a tight constraint when $\tilde{\chi}\gg 1$. We remark that  Eq.\ (\ref{largechi}) (strictly valid at $\tilde{\chi}\gg 1$, but possibly qualitatively correct also at  $\tilde{\chi}\simeq 1$) predicts strong effects  like those of Fig.\ \ref{fig2} down to  $\tilde{\chi}\simeq 1$, when this extra constraint disappears.     
To avoid a nonzero $V$, ${\bf A}$ can be generated alternatively by a pair of wires at $x=0$, but  different distances $d_1,d_2 \simeq a$ from the graphene plane (see Fig.\ \ref{fig1}). Opposite currents $I$ and $-I$ through such wires create a vector potential 
that falls off as $1/x^2$ at  $x\gg a$  and that has integral $ {\tilde{\chi}}=\int dx \,A_{{\rm y}}(x)= \mu_0 I(d_1-d_2)$. In our limit $r_{\rm s} \ll 1$, but $l\gg 1$ the deviations of this potential from the square-well shape Eq.\ (\ref{vec}) have negligible effects.

{\em Conclusions:} We have shown that the Coulomb repulsion between electrons can have  profound and unconventional effects on transport  in intrinsic graphene. General considerations predict singular transport across energy-independent 1D defects without Hartree potential. In addition we have shown that   interactions in graphene can enhance transport through static scatterers, in surprising contrast with the familiar result for 1D conductors. We have presented detailed calculations for vector potential scatterers that are induced by strain.  There the above effects have  exponentially large experimental signatures. 

The author thanks  Yu. V. Nazarov for  discussion   and the KITP at   UCSB for hospitality, where this work was supported in part by the NSF, Grant No. PHY05-51164.

 \end{document}